# Temporal Switching to Extend the Bandwidth of Thin Absorbers

Huanan Li,[1] and Andrea Alù[1,2,*]

[1]*Photonics Initiative, Advanced Science Research Center, City University of New York, New York, New York 10031, USA*

[2]*Physics Program, Graduate Center, City University of New York, New York, New York 10016, USA*

*Wave absorption in time-invariant, passive thin films is fundamentally limited by a trade-off between bandwidth and overall thickness. In this work, we investigate the use of temporal switching to reduce signal reflections from a thin grounded slab over broader bandwidths. We extend quasi-normal mode theory to time switching, developing an ab initio formalism that can model a broad class of time-switched structures. Our formalism provides optimal switching strategies to maximize the bandwidth over which minimal reflection is achieved, showing promising prospects for time-switched nanophotonic and metamaterial systems to overcome the limits of time-invariant, passive structures.*

Time-varying media have been recently raising significant attention in the broad physics and engineering communities, given their opportunities in the context of magnet-free non-reciprocity [1]-[5], multifunctional metasurfaces [6], PT-symmetric structures [7]-[8], amplification [9] and symmetry breaking for emission and absorption [10]. Since the pioneering work of Morgenthaler [11], research on dynamic media has mainly focused on time-periodic systems [12]-[17], for which the analysis is facilitated by the Floquet theorem. Imparting arbitrary time modulation schemes, beyond periodic, to wave-matter interactions can provide new possibilities and significantly broaden the field of dynamic media. Along this line, exotic wave phenomena have been recently shown in non-periodic time-varying structures, such as unlimited accumulation of energy [18]-[19] and arbitrary transport tuning via reconfigurable effective static potentials [20]. These studies have so far heavily relied on inverse engineering of the temporal modulation schemes, enabling only targeted functionalities and/or requiring simplifying assumptions like weak couplings. At the other extreme compared to periodically varying media, instantaneous temporal switching has drawn increasing attention as a platform for wave engineering [16],[21]-[22]. Abruptly changing in time the properties of an unbounded medium introduces a temporal boundary for wave propagation, dual of a spatial interface, generating forward and backward waves for which, instead of frequency, momentum is conserved and for which analytical solutions exist [22]. When considering more complicated structures, finite in one or more spatial dimensions and open to radiation, however, temporal switching interplays with other scattering phenomena and the analysis rapidly becomes very challenging in the absence of general constrains such as flux conservation.

[*]Corresponding author: aalu@gc.cuny.edu

To establish a thorough analysis of temporal switching in open systems, input-output techniques appear to be an ideal tool, since they can treat the field redistribution in a scatterer in time domain after the abrupt change of its properties as initial conditions to study the field evolution and associated scattering processes. This approach, initially established in quantum optics [24]-[25], has been applied to quantum networks [26]-[28], photon and charge transport [29]-[32] and quantum scattering [33]. As a variant for classical wave physics, coupled-mode theory (CMT) has been extensively used to model coupled resonator systems, well suited in regimes near resonance [34]-[36]. Except for few specific scenarios [37]-[38], its derivation follows from general principles, leaving many free phenomenological parameters undetermined. Nevertheless, there are two *ab initio* routes towards a rigorous input-output formalism. The first one is based on Feshbach's projector technique [39]-[40], based on the derivation of suitable system-bath Hamiltonians [33],[41]-[44], whereas the second one involves an expansion in terms of quasi-normal modes (QNMs) [45]-[50]. The former, involving Hermitian system Hamiltonian, cannot easily incorporate material loss and dispersion, while the latter typically focuses on scattering in frequency domain and for convenience deals with background and scattered fields, and thus its connection with CMT is vague since, for example, the background fields cannot be identified as inputs in CMT [48].

The goal of this work is twofold: first we extend the QNM approach to time-switched open systems [48]-[49], developing a generalized temporal input-output theory that bridges the gap between QNMs and conventional CMT. This approach is ideally suited for time-switched open systems, and yields new physical insights into the exotic wave phenomena arising in them. We then apply this theory to the analysis of temporal switching in thin-layer absorbers, in order to overcome the trade-off between bandwidth and thickness that applies to any time-invariant passive absorber [23].

***Thin absorbers and bandwidth limitations*** — Absorbing incident radiation over a large bandwidth is an important functionality for a variety of technologies, from radar technologies to anechoic measurements. Broadband absorption comes at the cost of thicker and heavier absorbing layers: consider for instance a Dallenbach screen, consisting of a homogeneous lossy nonmagnetic layer of thickness $d$ backed by a perfect electric conductor (PEC), as in Fig. **1**a. For normal incidence, the complex reflection coefficient $\rho(\omega) = \frac{\tanh(jn\omega d/c_0)-n}{\tanh(jn\omega d/c_0)+n}$ [51], where the refractive index $n = \sqrt{\varepsilon_r}\sqrt{1-j\sigma/(\varepsilon_0\varepsilon_r\omega)}$ is written in terms of the dielectric conductivity $\sigma$ and the real-valued (dispersion-less) relative permittivity $\varepsilon_r$, and $c_0 = 1/\sqrt{\mu_0\epsilon_0}$ is the speed of light in vacuum, with $\mu_0$ ($\epsilon_0$) being the permeability (permittivity). A thin screen with $\varepsilon_r \gg 1$ optimally absorbs for $\varepsilon_r^{opt} \approx \left(\frac{\lambda_0}{4d}\right)^2$ and $\hat{\sigma}^{opt} \approx 2\sqrt{\varepsilon_r^{opt}}\tanh^{-1}\left(1/\sqrt{\varepsilon_r^{opt}}\right)$, where $\hat{\sigma} = \sigma\eta_0 d$ with $\eta_0 = \sqrt{\mu_0/\epsilon_0}$ being the free-space impedance and $\lambda_0 = 2\pi c_0/\omega_0$ the free-space wavelength at the operating frequency $\omega_0$, yielding $\rho(\omega_0) = 0$. Correspondingly, the maximum bandwidth to least thickness ratio $\Delta\lambda/d \approx 32\rho_0/\pi$ for this geometry [52], where $\rho_0$ is the largest acceptable value of $|\rho(\omega)|$ within the operating bandwidth $\Delta\lambda$ centered around $\lambda_0$. This optimal condition falls within the more general Rozanov bound $\Delta\lambda/d < 16/|\ln\rho_0|$, valid for any nonmagnetic, passive, time-invariant absorber [23]. To reach this limit, further dispersion engineering of the material is generally required.

*QNM formulation*— Given its open resonance, the response of the Dallenbach screen in Fig. **1**(a) can be well captured by a QNM formulation. Assuming normal incidence along the positive *z*-direction, we can formulate the problem as

$$-j\frac{\partial}{\partial t}|\Psi(z,t)\rangle = \widehat{H}|\Psi(z,t)\rangle, \widehat{H} = \begin{pmatrix} j\sigma/(\varepsilon_0\varepsilon_r) & \hat{p}/\varepsilon_r \\ \hat{p} & 0 \end{pmatrix}, \quad (1)$$

where the differential operator $\hat{p} = jc_0\frac{\partial}{\partial z}$ and, without loss of generality, the state vector $|\Psi(z,t)\rangle = \begin{pmatrix} E_x \\ \eta_0 H_y \end{pmatrix}$ consists of the electric and magnetic field components $E_x$ and $H_y$ within the layer $-d = z_0 < z < 0$. The incident radiation is generally expressed as $|\Psi^{in}(z \leq z_0, t)\rangle \equiv a\left(t - \frac{z}{c_0}\right)\begin{pmatrix} 1 \\ 1 \end{pmatrix}$, and $|\Psi^{in}(z_0, t)\rangle$ determines the inhomogeneous incoming-wave boundary condition for Eq. (1) at all instants *t*.

We expand the scattered field $|\Psi_s(z,t)\rangle = \sum_n A_n(t) |\widetilde{\Psi}_n(z)\rangle$ [with the expansion coefficients $A_n(t)$] in terms of QNMs $|\widetilde{\Psi}_n(z,t)\rangle = |\widetilde{\Psi}_n(z)\rangle e^{j\widetilde{\omega}_n t}$, which individually satisfy $\widetilde{\omega}_n|\widetilde{\Psi}_n(z)\rangle = \widehat{H}|\widetilde{\Psi}_n(z)\rangle$ and the outgoing and PEC boundary conditions at $z = z_0$ and $z = 0$, yielding $\widehat{H}|\Psi_s(z,t)\rangle = \sum_n A_n(t)\,\widetilde{\omega}_n|\widetilde{\Psi}_n(z)\rangle$ [49],[53],[54]. $|\Psi_s(z,t)\rangle$ is obtained by writing the field $|\Psi(z,t)\rangle$ inside the layer as the superposition $|\Psi(z,t)\rangle = |\Psi_b(z,t)\rangle + |\Psi_s(z,t)\rangle$, where the background field $|\Psi_b(z,t)\rangle = a\left(t - \frac{z}{c_0}\right)\binom{1}{1} + a\left(t + \frac{z}{c_0}\right)\binom{-1}{1}$, $z_0 \leq z \leq 0$. Consequently, using Eq. (1) the scattered field $|\Psi_s(z,t)\rangle$ satisfies

$$-j\frac{\partial}{\partial t}|\Psi_s(z,t)\rangle = \widehat{H}|\Psi_s(z,t)\rangle + |s^+(z,t)\rangle, \tag{2}$$

where the inhomogeneous term $|s^+(z,t)\rangle = \begin{pmatrix} \frac{j\sigma}{\varepsilon_0 \varepsilon_r} & \left(\frac{1}{\varepsilon_r}-1\right)\hat{p} \\ 0 & 0 \end{pmatrix} |\Psi_b\rangle$. The reflected wave $|\Psi^{out}(z \leq z_0, t)\rangle = b\left(t + \frac{z}{c_0}\right)\binom{-1}{1}$ can be then obtained from the continuity of $|\Psi(z_0, t)\rangle$ at $z_0$, leading to

$$|\Psi^{out}(z_0, t)\rangle = a\left(t + \frac{z_0}{c_0}\right)\binom{-1}{1} + |\Psi_s(z_0, t)\rangle. \tag{3}$$

Our QNMs constitute a complete basis [55]-[56], and satisfy the orthonormality condition $\int_{z_0}^{0} \langle \widetilde{\Psi}_m^*(z)|M|\widetilde{\Psi}_n(z)\rangle dz = \delta_{mn}$, $M = \varepsilon_0 \begin{pmatrix} -\varepsilon_r & 0 \\ 0 & 1 \end{pmatrix}$, where $\langle \widetilde{\Psi}_m^*(z)| = |\widetilde{\Psi}_m(z)\rangle^T$ with the superscript T indicating the transpose operation. Explicitly, the QNMs of our geometry are

$$|\widetilde{\Psi}_n(z)\rangle = \begin{cases} c_n[\sin(\tilde{k}_n z), \; -\tan(\tilde{k}_n z_0)\cos(\tilde{k}_n z)]^T, z_0 \leq z \leq 0 \\ -c_n \sin(\tilde{k}_n z_0) e^{j\frac{\widetilde{\omega}_n}{c_0}(z-z_0)} [-1, \; 1]^T, z \leq z_0 \end{cases}, \tag{4}$$

where the complex wavenumbers $\tilde{k}_n = \frac{\sqrt{\varepsilon_r}}{c_0} \tilde{\omega}_n \sqrt{1 - j\sigma/(\varepsilon_0 \varepsilon_r \tilde{\omega}_n)}$ are written in terms of complex (angular) frequencies $\tilde{\omega}_n$ that satisfy the transcendental equation $\tan(\tilde{k}_n z_0) = -jc_0 \tilde{k}_n/\tilde{\omega}_n$, and the normalization factor $c_n = \sqrt{\frac{2}{d\varepsilon_0}} \left\{ \tan^2(\tilde{k}_n d) - \varepsilon_r + \frac{\sin(2\tilde{k}_n d)}{2\tilde{k}_n d} [\tan^2(\tilde{k}_n d) + \varepsilon_r] \right\}^{-1/2}$. These properties, combined with Eq. (2), provide a relation for the expansion coefficients $A_n(t) = \int_{z_0}^{0} \langle \widetilde{\Psi}_n^*(z)|M|\Psi_s(z,t)\rangle dz$ of the scattered field:

$$-j\frac{d}{dt} A_n(t) = \tilde{\omega}_n A_n(t) + \int_{z_0}^{0} \langle \widetilde{\Psi}_n^*(z)|M|s^+(z,t)\rangle dz. \qquad (5)$$

We can then derive the outgoing field $|\Psi^{out}(z_0, t)\rangle = b\left(t + \frac{z_0}{c_0}\right)\begin{pmatrix}-1\\1\end{pmatrix}$ using Eqs. (3) and (4):

$$b\left(t + \frac{z_0}{c_0}\right) = a\left(t + \frac{z_0}{c_0}\right) - \sum_n A_n(t)\, c_n \sin(\tilde{k}_n z_0). \qquad (6)$$

***Input-output formalism*** —Based on Eqs. (5) and (6), we can introduce an input-output formalism in terms of the modal coefficients $\psi_n(t) \equiv \int_{z_0}^{0} \langle \widetilde{\Psi}_n^*(z)|M|\Psi(z,t)\rangle dz$ of $|\Psi(z,t)\rangle$ with respect to the QNMs. Specifically, using the expression $A_n(t) = \psi_n(t) - \int_{z_0}^{0} \langle \widetilde{\Psi}_n^*(z)|M|\Psi_b(z,t)\rangle dz$ in Eqs. (5) and (6), we get

$$-j\frac{d}{dt}\psi_n(t) = \tilde{\omega}_n \psi_n(t) + 2jc_0\varepsilon_0 c_n \sin(\tilde{k}_n z_0)\, a^{in}(z_0, t),$$

$$b^{out}(z_0, t) = \int_{\left(t+\frac{2z_0}{c_0}\right)}^{t^-} K(t - t')a^{in}(z_0, t')dt' - \sum_n c_n \sin(\tilde{k}_n z_0)\, \psi_n(t), \qquad (7)$$

where $a^{in}(z_0, t) \equiv a\left(t - \frac{z_0}{c_0}\right) = \frac{1}{2\pi}\int_{-\infty}^{\infty} a(\omega)e^{j\omega(t-z_0/c_0)}d\omega$ and $b^{out}(z_0, t) \equiv b\left(t + \frac{z_0}{c_0}\right) = \frac{1}{2\pi}\int_{-\infty}^{\infty} b(\omega)e^{j\omega(t+z_0/c_0)}d\omega$ specify the incoming and outgoing fields at $z_0$ and time $t$, i.e., $|\Psi^{in}(z_0,t)\rangle = a^{in}(z_0,t)\binom{1}{1}$ and $|\Psi^{out}(z_0,t)\rangle = b^{out}(z_0,t)\binom{-1}{1}$, the integral kernel $K(\tau) = \delta\left(\tau + 2\frac{z_0}{c_0}\right) - \frac{c_0}{z_0}\sum_n \widehat{K}_n(\tau)$ with $\widehat{K}_n(\tau) \equiv c_n^2 \varepsilon_0 z_0 \sin(\tilde{k}_n z_0)\cos[\tilde{k}_n(c_0\tau + z_0)]\{\tan(\tilde{k}_n z_0) + \varepsilon_r \tan[\tilde{k}_n(c_0\tau + z_0)]\}$, and the integration range $\left[\left(t + \frac{2z_0}{c_0}\right)^-, t\right]$ covers the lower limit $t + \frac{2z_0}{c_0}$, ensuring that the Dirac delta $\delta\left(\tau + 2\frac{z_0}{c_0}\right)$ within $K(\tau)$ produces the expected term $a\left(t + \frac{z_0}{c_0}\right)$ [Eq. (6)]. We emphasize that the infinite summation in $K(\tau)$ typically does not converge, but the observable quantities, such as $b^{out}(z_0, t)$, do [33]. The *ab initio* input-output formulation Eq. (7) manifests causality, and can be applied beyond the weak-coupling limit, in contrast with the conventional CMT. To the best of our knowledge, Eq. (7) is the first example of rigorous input-output formalism based on QNMs, ideally suited to tackle time-switched open resonators as discussed later.

Our input-output formulation Eq. (7) collapses to conventional CMT in the limit in which the response is dominated by a single QNM-pair with complex-frequency $\widetilde{\omega}_{\pm 1}$, such that the internal fields $|\Psi(z,t)\rangle \approx \sum_{n=\pm 1}\psi_n(t)|\widetilde{\Psi}_n(z)\rangle$. This approximation is valid for our thin absorber when $\varepsilon_r \gg 1$ and the operating frequency $\omega_0 \sim \text{Re}(\widetilde{\omega}_{-1}) > 0$. In this regime, we find [57]

$$-j\frac{d\psi_{-1}(t)}{dt} = \widetilde{\omega}_{-1}\psi_{-1}(t) + jc_{-1}\sin(\tilde{k}_{-1}z_0)\sqrt{\frac{2}{\eta_0}}a_+^{in}(z_0,t),$$

$$b_+^{out}(z_0,t) = Ca_+^{in}(z_0,t) - c_{-1}\sin(\tilde{k}_{-1}z_0)\sqrt{\frac{2}{\eta_0}}\psi_{-1}(t),$$

(8)

where the complex wave amplitudes $a_+^{in}(z_0, t) = \sqrt{2/\eta_0} \frac{1}{2\pi} \int_0^\infty a(\omega) e^{j\omega(t-z_0/c_0)} d\omega$ and $b_+^{out}(z_0, t) = \sqrt{2/\eta_0} \frac{1}{2\pi} \int_0^\infty b(\omega) e^{j\omega(t+z_0/c_0)} d\omega$ are normalized to the time-averaged power flux under the slowly-varying envelope approximation. The direct coupling coefficient C connecting the complex excitation with the outgoing fields is $C \approx e^{2j\omega_0 z_0/c_0} + \int_{-2z_0/c_0}^{0} \widehat{K}_{-1}(\tau) e^{-j\tau\omega_0} c_0/z_0 d\tau$, with $C \to 1$ for our thin absorber in the limit $\varepsilon_r \gg 1$. In Ref. [57], we examine Eq. (8) analytically in the lossless scenario and compare it with the general principle formulation. As expected, Eq. (8) converges to the conventional CMT [36] in the weak-coupling regime. Yet, when strong coupling and/or material dissipation is present, for which CMT cannot apply, our *ab initio* input-output formalism Eq. (7) provides a generalized tool to analytically study the scattering problem.

We apply our formalism to study the scattering from a time-switched Dallenbach screen. Consider, Fig. **1**(a), a wave packet $a\left(t - \frac{z}{c_0}\right) = f\left(t - \frac{z-z_i}{c_0}\right) \cos\left[\omega_0 \left(t - \frac{z-z_i}{c_0}\right)\right]$ with normalized envelope $f(t) = a_{peak} e^{-2t^2 \ln 2/\delta t^2}$, where $z_i \ll z_0$ determines the initial position of the Gaussian pulse with peak value $a_{peak} = \left(\frac{2\sqrt{\ln 2}}{\sqrt{\pi}\delta t}\right)^{1/2}$. In frequency domain, the incident wave packet $a(\omega) = \frac{1}{2} e^{j\omega z_i/c_0}[\tilde{f}(\omega - \omega_0) + \tilde{f}(\omega + \omega_0)]$, where $\tilde{f}(\omega) = 2\left(\frac{\sqrt{\pi \ln 2}}{\delta\omega}\right)^{1/2} e^{-2\omega^2 \ln 2/\delta\omega^2}$ with $\delta\omega = 4\ln 2/\delta t$ being the FWHM. We readily calculate the reflection $b^{out}(z_0, t)$ and time-dependent internal fields $|\Psi(z,t)\rangle, z_0 \leq z \leq 0$:

$$b^{out}(z_0, t) = \frac{1}{2\pi} \int_{-\infty}^{\infty} \left[-a(\omega) e^{-\frac{j\omega z_0}{c_0}}\right] \rho(\omega) e^{j\omega t} d\omega,$$

$$|\Psi(z,t)\rangle = \frac{1}{2\pi} \int_{-\infty}^{\infty} \frac{2a(\omega) e^{j\omega(t-z_0/c_0)}}{\sin(\tilde{k}z_0) + jn\cos(\tilde{k}z_0)} \begin{pmatrix} \sin(\tilde{k}z) \\ jn\cos(\tilde{k}z) \end{pmatrix} d\omega,$$

(9)

where $\rho(\omega)$ is the complex reflection coefficient derived above, and the complex wave number $\tilde{k} \equiv \frac{\omega n}{c_0}$. We can employ Eq. (7) to evaluate the reflection $b^{out}(z_0, t)$ starting from an arbitrary time $t_s$, after knowing the initial condition $\psi_n(t_s) = \int_{z_0}^{0} \langle \widetilde{\Psi}_n^*(z) | M | \Psi(z, t_s) \rangle dz$. Note that $\psi_n(0) = 0$, since initially the incoming wave packet is far away from the screen, and $\psi_n(t_s), t_s > 0$ can be calculated exactly from the knowledge of the internal field $|\Psi(z, t_s)\rangle$ in Eq. (9) [58].

Consider the Dallenbach screen with $d/\lambda_0 = 0.03$ and optimal material parameters $\varepsilon_r^{opt} = 69.85$ and $\hat{\sigma}^{opt} = 1.998$. In Fig. 1(b), we show the reflection spectrum (green-solid line), together with the frequency spectrum $|a(\omega)/a(\omega_0)|$ (red-dashed line) for an impinging pulse with FWHM $\delta\omega = 0.2\omega_0$ and initial position $z_i = 200 z_0$. In Fig. **1**(c), we show the evolution of the first few (normalized) QNM frequencies $\hat{y}_n = \widetilde{\omega}_n z_0 \sqrt{\varepsilon_r}/c_0, n = \pm 1, \pm 2$ as we vary $\varepsilon_r$ from 140 to 2, with $\hat{\sigma} = \hat{\sigma}^{opt}$. The green plus and red cross symbols correspond respectively to $\varepsilon_r = \varepsilon_r^{opt}$ for the optimal absorber and $\varepsilon_r = \varepsilon_r^{EP} \approx 2.1$ for the permittivity value corresponding to an exceptional point (EP) where the $n = \pm 1$ QNMs coalesce [57]. Around the EP, the complex frequency $\hat{y}_{\pm 1}(\varepsilon_r)$ of the $n = \pm 1$ QNMs follow a square root behavior, a characteristic signature of second-order EP singularities, as $\hat{y}_{\pm 1}(\varepsilon_r) \approx -j1.84 \pm 1.53\sqrt{\varepsilon_r - \varepsilon_r^{EP}}$ [57]. Using Eq. (7) and the calculated QNMs, we evaluate the temporal evolution $b^{out}(z_0, t)/a_{peak}$ of the slab reflection [green circles in Fig. **1**(d)], which fits perfectly with the exact result provided by Eq. (9) [black solid line]. In the same subfigure, we also show the result from conventional CMT Eq. (8), which works well in the $\varepsilon_r \gg 1$ regime except for an initial short period.

***Time-switched absorber*** — Our ab initio formalism is ideally suited to study the effect of abruptly switching the properties of the Dallenbach screen. Specifically, we consider the case in which the relative permittivity abruptly changes at time $t_s$ from $\varepsilon_1$ to $\varepsilon_2$. The time-domain response of the screen after $t_s$ cannot be deduced from Fourier analysis, but it can be readily obtained from our input-output formalism Eq. (7), once knowing the internal field $|\Psi(z, t_s^+)\rangle, z_0 \leq z \leq 0$ immediately after the switch. To this end, we employ the continuity conditions for the electric displacement and magnetic induction across the temporal boundary at $t_s$ [59]-[60], which equivalently reads $|\Psi(z, t_s^+)\rangle = \begin{pmatrix} \varepsilon_1/\varepsilon_2 & 0 \\ 0 & 1 \end{pmatrix} |\Psi(z, t_s^-)\rangle$. In turn, the internal field $|\Psi(z, t_s^-)\rangle$ immediately before the switch can be calculated from Eq. (9). Generally, the abrupt change of material properties does not preserve energy, and the injected energy $\Delta E \approx |\psi_{-1}(t_s^+)|^2 - |\psi_{-1}(t_s^-)|^2$ using Eq. (8).

In Fig. 2, we study the effect of this abrupt switching on the reflection coefficient of the optimal absorber. In Fig. 2(a), we abruptly change $\varepsilon_r$ at an arbitrary switching time $t_s \in [45, 55]$ (in units of $1/\omega_0$) from $\varepsilon_1 = \varepsilon_r^{opt}$ to $\varepsilon_2 \in [10, 120]$. For each pair $(t_s, \varepsilon_2)$, we calculate the relative total reflected energy $10\log_{10}|E^{out}/E^{in}|$, where $E^{out/in} = \int_0^\infty P^{out/in}(z_0, t)dt$ with $P^{out} = \frac{1}{\eta_0}(b^{out})^2$ and $P^{in} = \frac{1}{\eta_0}(a^{in})^2$ being the total outgoing/incoming energy. The total reflected energy (upper orange surface) is larger than the optimal static absorber, i.e., the case $\varepsilon_2 = \varepsilon_1 = \varepsilon_r^{opt}$. However, the switching operation spreads the reflected energy over a wide range of frequencies outside the bandwidth of the incident pulse. The relative reflected energy, as well as the visibility of the target, is therefore reduced (lower green mesh surface) when we replace the total reflected energy $E^{out}$ with the filtered energy $E_{bw}^{out}$ within the pulse frequency window $(\omega_0 - \Delta\omega, \omega_0 + \Delta\omega)$ with $\Delta\omega = \delta\omega = 0.2\omega_0$, which covers 98% of the total incoming energy. The switching mechanism effectively minimizes detectability in reflection by spreading the non-absorbed energy over a broad frequency range, as explored in Refs. [61]-[65] for lossless periodically modulated screens. Our approach combines reduced reflection due to absorption with spreading arising through a single switching event, making the functionality efficient and effective. We envision these devices being operated passively as a regular absorber using phase change materials, with its phase transition being triggered by the arrival of an incoming pulse to improve their bandwidth performance beyond the limits of passive absorbers.

For given switching time $t_s$, the filtered reflection energy $E_{bw}^{out}$ monotonically decreases with increasing switching strength $|\varepsilon_2 - \varepsilon_1|$ when the initial permittivity $\varepsilon_1 = \varepsilon_r^{opt}$, until the system reaches the EP in Fig. 1(c), see Fig. 2(b) plotted for $t_s = 49\ [1/\omega_0]$. Around the EP, our input-output formalism converges increasingly slow, as also seen from the deteriorated comparison with conventional CMT Eq. (8) [blue dashed line]. In the left inset, we zoom around $\varepsilon_2 \approx \varepsilon_r^{EP}$, highlighting the rapid change in reflection, and thus the sensitivity around the EP, with a deterioration in overall response. In the right inset, we plot the normalized injected energy $\Delta E/E^{in}$ by switching as a function of $\varepsilon_2$. Depending on whether $\varepsilon_2 < \varepsilon_1 = \varepsilon_r^{opt}$, we inject or extract increasing energy as the switching amplitude $|\varepsilon_1 - \varepsilon_2|$ grows.

For given switching amplitude, we find an optimal switching time $t_s^{opt}$ that minimizes $E_{bw}^{out}$. In Fig. 3, we fix $\varepsilon_2 = 8$, and calculate the effective reflection spectrum $|\rho| = |\tilde{b}^{out}(\omega)/a(\omega)|$ for various impinging pulses with FWHM $\delta\omega/\omega_0 = 0.05, 0.1, 0.15, 0.2$. For each scenario, we employ $t_s^{opt}$, which occurs right after the peak of each impinging pulse passes through the front surface of the absorber, and can be easily implemented by triggering the switch when the input energy decays. The reflected energy is spread over broader bandwidths, well beyond the optimal reflection achievable in a passive scenario in Fig. 1(b). The bandwidth can be further increased by adjusting the desired minimum attainable reflection level, or by adding a second switching for longer pulses.

*Conclusions*— In this work, we have analyzed the effect of abrupt time switching on the overall bandwidth of thin absorbers, showing that the overall bandwidth of reflection can be largely enhanced compared to passive time-invariant slabs. Our analysis is based on a newly introduced input-output formalism that efficiently captures the scattering phenomena of open systems, ideally suited to analyze temporally switched scenarios. This general formulation, derived from a QNM expansion, is exact and enables a clean first-principle derivation of conventional CMT. Using this approach, we found that temporal switching can effectively reduce the reflection of impinging pulses, and may strengthen the effectiveness of non-magnetic thin absorbers. Our approach can be extended to time switched metamaterials and metasurfaces, and scattering problems involving nanoparticles, for which the QNM expansion can be efficiently employed.

*Acknowledgements*— This work was supported by the Air Force Office of Scientific Research and the Simons Foundation.

**References**

[1] Z. Yu and S. Fan, "Complete optical isolation created by indirect interband photonic transitions," Nat. Photonics **3**, 91 (2009).

[2] D. L. Sounas and A. Alù, "Angular-momentum-biased nanorings to realize magnetic-free integrated optical isolation," ACS Photon. **1**, 198 (2014).

[3] S. Taravati, N. Chamanara, and C. Caloz, "Nonreciprocal electromagnetic scattering from a periodically space-time modulated slab and application to a quasisonic isolator," Phys. Rev. B **96**, 165144 (2017).

[4] Y. Shi, S. Han, and S. Fan, "Optical circulation and isolation based on indirect photonic transitions of guided resonance modes," ACS Photonics **4**, 1639 (2017).

[5] N. A. Estep, D. L. Sounas, J. Soric and A. Alù, "Magnetic-free non-reciprocity and isolation based on parametrically modulated coupled-resonator loops," Nat. Phys. **10**, 923 (2014).

[6] X. Wang, A. Diaz-Rubio, H. Li, S. A. Tretyakov, and A. Alù, "Theory and design of multifunctional space-time metasurfaces," Phys. Rev. Appl., **13**, 044040 (2020).

[7] R. El-Ganainy, K. G. Makris, and D. N. Christodoulides, "Local PT invariance and supersymmetric parametric oscillators," Phys. Rev. A **86**, 033813 (2012).

[8] T. T. Koutserimpas, A. Alù, and R. Fleury, "Parametric amplification and bidirectional invisibility in PT-symmetric time-Floquet systems," Phys. Rev. A **97**, 013839 (2018).

[9] T. T. Koutserimpas and R. Fleury, "Nonreciprocal gain in non-hermitian time-Floquet systems," Phys. Rev. Lett., **120**, 087401 (2018).

[10] Y. Hadad, J. Soric, and A. Alù, "Breaking temporal symmetries for emission and absorption," Proc. Natl. Acad. Sci. USA **113**, 3471 (2016).

[11] F. R. Morgenthaler, "Velocity Modulation of Electromagnetic Waves," *IRE Trans. Microw. Theory Tech.*, **6**, 167 (1958).

[12] D. E. Holberg and K. S. Kunz, "Parametric properties of fields in a slab of time-varying permittivity," IEEE Transactions on Antennas and Propagation, **14**, 183 (1966).

[13] J. R. Zurita-Sánchez, P. Halevi, and J. C. Cervantes- González, "Reflection and transmission of a wave incident on a slab with a time-periodic dielectric function ε(t)," Phys. Rev. A **79**, 053821 (2009).

[14] J. S. Martínez-Romero, O. M. Becerra-Fuentes, and P. Halevi, "Temporal photonic crystals with modulations of both permittivity and permeability," Phys. Rev. A **93**, 063813 (2016).

[15] N. Wang, Z-Q Zhang, and C. T. Chan, "Photonic Floquet media with a complex time-periodic permittivity," Phys. Rev. B **98**, 085142 (2018).

[16] T. T. Koutserimpas and R. Fleury, "Electromagnetic Waves in a Time Periodic Medium With Step-Varying Refractive Index," IEEE Trans. Antennas Propag., **66**, 5300 (2018).

[17] P. A. Pantazopoulos and N. Stefanou, "Layered optomagnonic structures: Time Floquet scattering-matrix approach," Phys. Rev. B, **99**, 144415 (2019).

[18] M. S. Mirmoosa, G. A. Ptitcyn, V. S. Asadchy, and S. A. Tretyakov, "Time-Varying Reactive Elements for Extreme Accumulation of Electromagnetic Energy," Phys. Rev. Appl., **11**, 014024 (2019).

[19] D. L. Sounas, "Virtual perfect absorption through modulation of the radiative decay rate," Phys. Rev. B **101**, 104303 (2020).

[20] H. Li, T. Kottos, and B. Shapiro, "Driving-induced metamorphosis of transport in arrays of coupled resonators," Physical Review A **97**, 023846 (2018).

[21] V. Pacheco-Peña and N. Engheta, "Antireflection temporal coatings," Optica **7**, 323 (2020).

[22] C. Caloz and Z.-L. Deck-Léger, "Spacetime metamaterials," arXiv:1905.00560 (2019).

[23] K. N. Rozanov, "Ultimate thickness to bandwidth ratio of radar absorbers," IEEE Trans. Antennas Propagat., **48**, 1230 (2000).

[24] M. J. Collett and C. W. Gardiner, "Squeezing of intracavity and traveling-wave light fields produced in parametric amplification," Phys. Rev. A **30**, 1386 (1984).

[25] C. W. Gardiner and M. J. Collett, "Input and output in damped quantum systems: Quantum stochastic differential equations and the master equation," Phys. Rev. A **31**, 3761 (1985).

[26] B. Yurke and J. S. Denker, "Quantum network theory," Phys. Rev. A **29**, 1419 (1984).

[27] J. Zhang, Y.-X. Liu, R.-B. Wu, K. Jacobs, and F. Nori, "Non-Markovian quantum input-output networks," Phys. Rev. A **87**, 032117 (2013).

[28] A. Reiserer and G. Rempe, "Cavity-based quantum networks with single atoms and optical photons," Rev. Mod. Phys. **87**, 1379 (2015).

[29] S. Xu and S. Fan, "Input-output formalism for few-photon transport: A systematic treatment beyond two photons," Phys. Rev. A **91**, 043845 (2015).

[30] T. Caneva, M. T. Manzoni, T. Shi, J. S. Douglas, J. I. Cirac, and D.E. Chang, "Quantum dynamics of propagating photons with strong interactions: a generalized input-output formalism," New J. Phys. **17**, 113001 (2015).

[31] J. Liu and D. Segal, "Generalized input-output method to quantum transport junctions. I. General formulation," Phys. Rev. B **101**, 155406 (2020).

[32] J. Liu and D. Segal, "Generalized input-output method to quantum transport junctions. II. Applications," Phys. Rev. B **101**, 155407 (2020).

[33] D. Lentrodt and J. Evers, "*Ab initio* few-mode theory for quantum potential scattering problems," Phys. Rev. X **10**, 011008 (2020).

[34] H. A. Haus, Waves and Fields in Optoelectronics (Prentice-Hall, Englewood Cliffs, NJ, 1984).

[35] S. Fan, W. Suh, and J.D. Joannopoulos, "Temporal coupled-mode theory for the fano resonance in optical resonators," J. Opt. Soc. Am. A **20**, 569 (2003).

[36] W. Suh, Z. Wang, and S. Fan, "Temporal coupled-mode theory and the presence of non-orthogonal modes in lossless multimode cavities," IEEE J. Quantum Electron. **40**, 1511 (2004).

[37] J. K. S. Poon, and A. Yariv, "Active coupled-resonator optical waveguides. I. Gain enhancement and noise," J. Opt. Soc. Am. B **24**, 2378 (2007).

[38] H. Li, A. Mekawy, A. Krasnok, and A. Alù, "Virtual Parity-Time Symmetry," Phys. Rev. Lett. **124**, 193901 (2020).

[39] H. Feshbach, "Unified theory of nuclear reactions," Ann. Phys. (N.Y.) **5**, 357 (1958).

[40] H. Feshbach, "A unified theory of nuclear reactions. II," Ann. Phys. (NY) **19**, 287 (1962).

[41] C. Viviescas and G. Hackenbroich, "Field quantization for open optical cavities," Phys. Rev. A **67**, 013805 (2003).

[42] C. Viviescas and G. Hackenbroich, "Quantum theory of multimode fields: applications to optical resonators," J. Opt. B **6**, 211 (2004).

[43] D. V. Savin, V. V. Sokolov, and H.-J. Sommers, "Is the concept of the non-Hermitian effective Hamiltonian relevant in the case of potential scattering," Phys. Rev. E **67**, 026215 (2003).

[44] W. Domcke, "Projection-operator approach to potential scattering," Phys. Rev. A **28**, 2777 (1983).

[45] S. M. Dutra and G. Nienhuis, "Quantized mode of a leaky cavity," Phys. Rev. A **62**, 063805 (2000).

[46] B. Vial, F. Zolla, A. Nicolet, and M. Commandré, "Quasimodal expansion of electromagnetic fields in open two-dimensional structures," Phys. Rev. A **89**, 023829 (2014).

[47] E. A. Muljarov and W. Langbein, "Resonant-state expansion of dispersive open optical systems: creating gold from sand," Phys. Rev. B **93**, 075417 (2016).

[48] P. Lalanne, W. Yan, K. Vynck, C. Sauvan, and J.-P. Hugonin, "Light Interaction with Photonic and Plasmonic Resonances," Laser Photonics Rev. **12**, 1700113 (2018).

[49] W. Yan, R. Faggiani, and P. Lalanne, "Rigorous modal analysis of plasmonic nanoresonators," Phys. Rev. B **97**, 205422 (2018).

[50] S. Franke, S. Hughes, M. K. Dezfouli, P. T. Kristensen, K. Busch, A. Knorr, and M. Richter, "Quantization of quasinormal modes for open cavities and plasmonic cavity quantum electrodynamics," Phys. Rev. Lett. **122**, 213901 (2019).

[51] D. M. Pozar, Microwave Engineering (Wiley, New York, 1998).

[52] G. Ruck, D. E. Barrick, W. D. Stuart, and C. K. Krichbaum, Radar Cross Section Handbook. (New York: Plenum, 1970).

[53] R. Haberman, Elementary Applied Partial Differential Equations with Fourier Series and Boundary Value Problems (Prentice-Hall, New Jersey, 1987).

[54] As a comparison, for a valid QNM expansion $|\Psi(z,t)\rangle = \sum_n \psi_n(t) |\widetilde{\Psi}_n(z)\rangle$ of the state vector $|\Psi(z,t)\rangle$, a seemingly similar equality $\hat{H}|\Psi(z,t)\rangle = \sum_n \psi_n(t)\, \widetilde{\omega}_n |\widetilde{\Psi}_n(z)\rangle$ does not hold, which hinders a construction of an input-output formalism using QNMs [see Eq. (7)] directly from Eq. (1).

[55] P. T. Leung, S. Y. Liu, and K. Young, "Completeness and time-independent perturbation of the quasinormal modes of an absorptive and leaky cavity," Physical Review A **49**, 3982 (1994).

[56] E. S. C. Ching, P. T. Leung, A. Maassen van den Brink, W. M. Suen, S. S. Tong, and K. Young, "Quasinormal-mode expansion for waves in open systems," Rev. Mod. Phys. **70**, 1545 (1998).

[57] See the supplement for further details.

[58] $\psi_n(t_s), t_s > 0$ can be also obtained by solving Eq. (7) with the initial condition $\psi_n(0) = 0$.

[59] Y. Xiao, D. N. Maywar, and G. P. Agrawal, "Reflection and transmission of electromagnetic fields from a temporal boundary," Opt. Lett. **39**, 574 (2014).

[60] A. G. Hayrapetyan, J. B. Götte, K. K. Grigoryan, S. Fritzsche, and R. G. Petrosyan, "Electromagnetic wave propagation in spatially homogeneous yet smoothly time-varying dielectric media," J. Quant. Spectrosc. Radiat. Transfer., **178**, 158 (2016).

[61] A. Tennant, "Reflection properties of a phase modulating planar screen," Electron. Lett. **33**, 1768 (1997).

[62] B. Chambers and A. Tennant, "General analysis of the phase-switched screen. Part 1: The single layer case," IEE Proc. Radar, Sonar Navigation **149**, 243 (2002).

[63] B. Chambers and A. Tennant, "The phase-switched screen," IEEE Antennas Propagation Mag. **46**, 23 (2004).

[64] B. Chambers, and A. Tennant, "A smart radar absorber based on the phase-switched screen," IEEE Trans Antennas Propag. **53**, 394 (2005).

[65] B. Chambers, and A. Tennant, "Active Dallenbach radar absorber," IEEE International Symposium on Antennas and Propagation, Albuquerque, New Mexico, USA (2006).

**Figures**

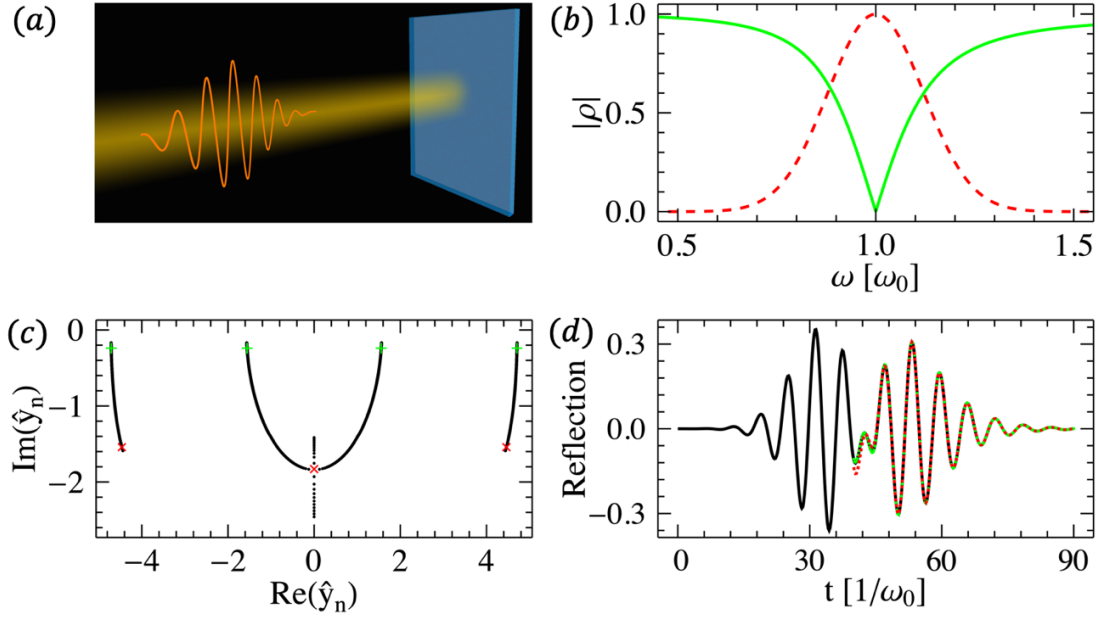

**Fig. 1.** ($a$) A Gaussian pulse impinges on the Dallenbach absorber. ($b$) Reflection magnitude $|\rho|$ versus incident (angular) frequency $\omega$ for the optimal thin absorber with $\varepsilon_r^{opt} \approx 69.85$, $\hat{\sigma}^{opt} \approx 1.998$ and $d/\lambda_0 = 0.03$. Red-dashed line represents the frequency spectrum $|a(\omega)/a(\omega_0)|$ of a pulse with FWHM $\delta\omega = 0.2\omega_0$. ($c$) Trajectories of the first two (normalized) QNM pairs $\hat{y}_n$, $n = \pm 1, \pm 2$ as $\varepsilon_r$ varies from 140 to 2 while $\hat{\sigma} = \hat{\sigma}^{opt}$. Two special scenarios when $\varepsilon_r = \varepsilon_r^{opt}, \varepsilon_r^{EP}(\approx 2.1)$ are marked respectively by green plus and red cross symbols. ($d$) Time-domain reflection $b^{out}(z_0, t)/a_{peak}$ [black solid line] for the impinging pulse located initially at $z_i = 200z_0$. The results calculated at time $t_s = 40[1/\omega_0]$ based on Eq. (7) [green circles] and Eq. (8) with $C = 1$ [red dashed line and see Ref. [57]] are also shown.

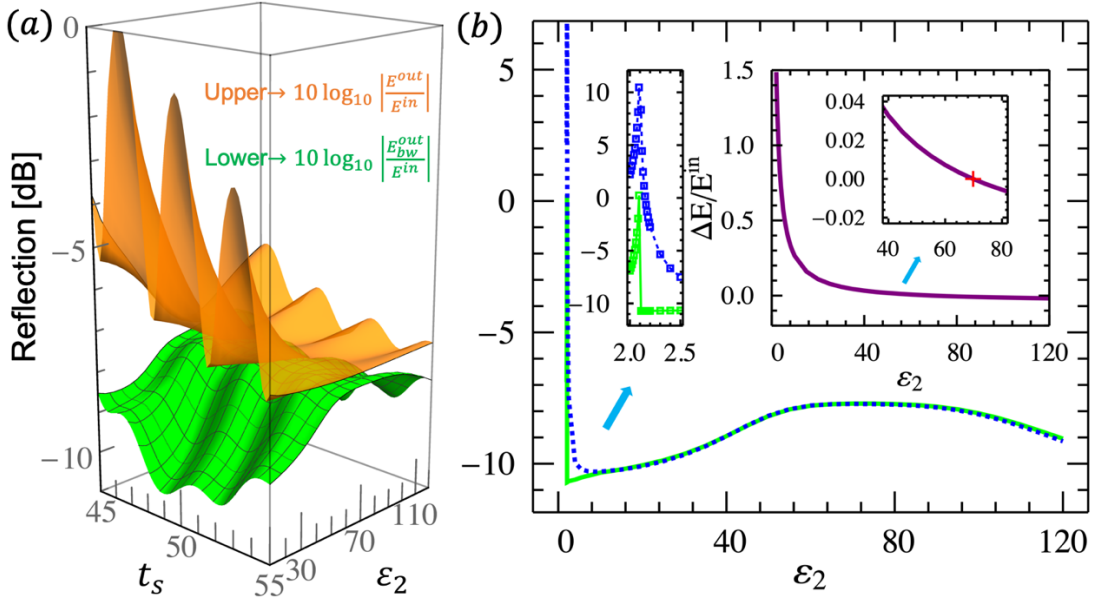

**Fig. 2.** Performance of switched absorbers. (a) The upper orange surface presents the total reflected energy $10\log_{10}|E^{out}/E^{in}|$ versus switching time $t_s\omega_0 \in [45, 55]$ and relative permittivity $\varepsilon_2 \in [10, 120]$ after switching. The lower green mesh surface shows the reflected energy $10\log_{10}|E_{bw}^{out}/E^{in}|$ within the bandwidth $(\omega_0 - \delta\omega, \omega_0 + \delta\omega)$ of the incoming pulse, with $\delta\omega = 0.2\omega_0$. (b) Reflected energy $10\log_{10}|E_{bw}^{out}/E^{in}|$ in (a) as a function of $\varepsilon_2 \in [2, 120]$ when $t_s = 49 \,[1/\omega_0]$. The result from Eq. (8) is also shown (blue dashed line). (Left inset) Zoom around the EP $\varepsilon_2 = \varepsilon_r^{EP}$. (Right inset) Normalized switching energy $\Delta E/E^{in}$ versus $\varepsilon_2$, which changes sign at $\varepsilon_2 = \varepsilon_r^{opt}$ [see inset for the portion around $(\varepsilon_r^{opt}, 0)$ (red plus symbol)]. Before switching, the screen operates in the optimal static mode, see Fig. **1**.

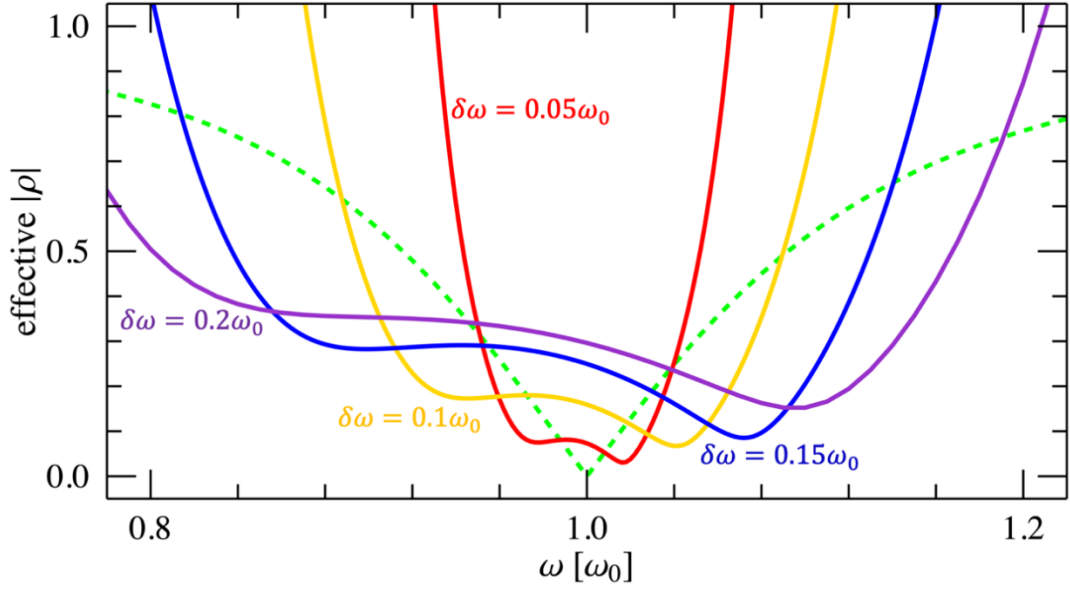

**Fig. 3.** The effective reflection spectrum $|\rho|$ of the time-switched absorber for various impinging Gaussian pulses of FWHM $\delta\omega/\omega_0 = 0.05, 0.1, 0.15, 0.2$, with sufficiently larger initial positions $z_i/z_0 = 750, 350, 300, 200$, respectively. Here, the absorber permittivity switches from $\varepsilon_1 = \varepsilon_r^{opt}$ to $\varepsilon_2 = 8$ at the optimal switching time $t_s^{opt} \approx 197, 90, 71, 49\ [1/\omega_0]$. The green-dashed line indicates the reflection spectrum of the optimal static absorber in Fig. **1**(b). Other parameters are the same as Fig. **1** and **2**.

# Supplementary material

## Temporal Switching to Extend the Bandwidth of Thin Absorbers

Huanan Li,[1] and Andrea Alù[1,2,*]

[1]*Photonics Initiative, Advanced Science Research Center, City University of New York, New York, New York 10031, USA*

[2]*Physics Program, Graduate Center, City University of New York, New York, New York 10016, USA*

*To whom correspondence should be addressed: aalu@gc.cuny.edu

## I. First-principle derivation of conventional coupled-mode theory (CMT)

In this section, we present a first-principle derivation of the conventional CMT using the input-output formulation Eq. (7) of the main text. We first note that the QNMs appear in pairs [66] and their complex wave characteristics obey $\widetilde{\omega}_{-n} = -\widetilde{\omega}_n^*$, $\widetilde{k}_{-n} = -\widetilde{k}_n^*$ and $(c_{-n})^2 = (c_n^*)^2$ with the label $n = \pm 1, \pm 2, \pm 3, \cdots$. Therefore, to ensure the real-valued nature of the outgoing radiation $|\Psi^{out}(z_0, t)\rangle$ and the internal fields $|\Psi(z, t)\rangle$, these QNM pairs $n = \pm 1, \pm 2, \pm 3, \cdots$ should be employed as units for the QNM expansion and thus the summation in Eq. (7) of the main text. A specific label scheme for the QNMs is fixed when setting $\widetilde{\omega}_n = \hat{y}_n c_0/(z_0 \sqrt{\varepsilon_r})$ and assuming $0 \leq \text{Re}(\hat{y}_1) < \text{Re}(\hat{y}_2) < \cdots$. As mentioned in the main text, we consider one QNM-pair approximation associated with the complex-frequency pair $\widetilde{\omega}_{\pm 1}$, such that the internal fields $|\Psi(z, t)\rangle \approx \sum_{n=\pm 1} \psi_n(t) |\widetilde{\Psi}_n(z)\rangle$. This approximation is valid for our thin absorber where $\varepsilon_r \gg 1$ and when the operating frequency $\omega_0 \sim \text{Re}(\widetilde{\omega}_{-1}) > 0$. As we know, the conventional CMT deals with the energy-normalized positive-frequency component of the mode amplitude [34]. Here, it can be identified to be simply $\psi_{-1}(t)$ since, in the weak coupling limit $\varepsilon_r \to \infty$ and the lossless scenario $\sigma = 0$, its complex frequency $\widetilde{\omega}_{-1}$ approaches to the (positive) resonant frequency and the total stored energy inside the layer, i.e., $E_{tot}(t) = \int_{z_0}^{0} \frac{1}{2} \varepsilon_0 \langle \Psi | \begin{pmatrix} \varepsilon_r & 0 \\ 0 & 1 \end{pmatrix} | \Psi \rangle dz$, is shared by the QNM pair $n = \pm 1$ as $|\psi_{-1}(t)|^2$. As for excitations, we follow the CMT convention to adopt the power-normalized complex amplitude $a_+^{in}(z_0, t)$ of positive frequencies, i.e., $a_+^{in}(z_0, t) = \sqrt{2/\eta_0} \frac{1}{2\pi} \int_0^\infty a(\omega) e^{j\omega(t-z_0/c_0)} d\omega$. Under the slow-envelope approximation, i.e., $a_+^{in}(z_0, t) = A_+^{in}(t) e^{j\omega_0 t}$ with $A_+^{in}(t + \Delta t) \approx A_+^{in}(t)$ when $\Delta t \sim O(2\pi/\omega_0)$, the time-averaged incoming power with respect to $2\pi/\omega_0$ is $\langle P^{in}(z_0, t) \rangle = |a_+^{in}(z_0, t)|^2$ as it should. For the outgoing radiation, we have the power-normalized complex amplitude $b_+^{out}(z_0, t) = \sqrt{2/\eta_0} \frac{1}{2\pi} \int_0^\infty b(\omega) e^{j\omega(t+z_0/c_0)} d\omega$, satisfying similarly $\langle P^{out}(z_0, t) \rangle = |b_+^{out}(z_0, t)|^2$. Finally, decoupling the positive frequency with the negative ones in Eq. (7) of the main text in the spirit of the rotating-wave approximation and employing the slow-envelope approximation for the complex excitation $a_+^{in}(z_0, t)$, we can reach the conventional CMT under one QNM-pair approximation, i.e., Eq. (8) of the main text.

Next, we consider lossless screens and analytically examine the derived conventional CMT. We start our discussion by rewriting Eq. (8) [see main text] in the conventional form [68]

$$\frac{d\psi_{-1}(t)}{dt} = (j\Omega - \Gamma)\psi_{-1}(t) + K^T a_+^{in}(z_0, t),$$
$$b_+^{out}(z_0, t) = C a_+^{in}(z_0, t) + D\psi_{-1}(t),$$
(S10)

where the resonant (angular) frequency of the single mode $\Omega = \text{Re}(\widetilde{\omega}_{-1})$, its decay rate $\Gamma = \text{Im}(\widetilde{\omega}_{-1})$, the coupling between the mode and the port $K = D = -c_{-1} \sin(\tilde{k}_{-1} z_0) \sqrt{\frac{2}{\eta_0}}$ and the direct path between the incoming and outgoing wave in the port $C \approx e^{2j\omega_0 z_0/c_0} +$ $\int_{-2z_0/c_0}^{0} \widehat{K}_{-1}(\tau)e^{-j\tau\omega_0}c_0/z_0 d\tau$. In the scenario when $\sigma = 0$ as focused here, we have the following relationships $\tilde{k}_n z_0 = \hat{y}_n$, $\tan\hat{y}_n = -j\sqrt{\varepsilon_r}$ and $c_n^2 \varepsilon_0 z_0 = 1/\varepsilon_r$. In addition, the (normalized) complex frequency $\hat{y}_n \equiv \widetilde{\omega}_n z_0 \sqrt{\varepsilon_r}/c_0$ of QNMs can be explicitly given as $-\hat{y}_{-n}^* = \hat{y}_n = \frac{(2n-1)\pi}{2} - j\text{Re}(\text{ArcTanh}\sqrt{\varepsilon_r})$, $n = 1,2,3,\cdots$. Furthermore, the propagator $\widehat{K}_n(\tau)$ due to the nth QNM for the direct path (see the main text) can be simplified greatly to be $\widehat{K}_n(\tau) = -\cos\left[\hat{y}_n\left(2 + \frac{c_0}{z_0}\tau\right)\right]$. Therefore, we can easily calculate the (approximated) direct path C in Eq. (S10), which turns out to be $C = 1 + j\left(\frac{8}{\pi} - \pi\right)\frac{1}{\sqrt{\varepsilon_r}} + O(1/\varepsilon_r)$ in the weak-coupling limit when $\varepsilon_r \to \infty$. We point out that in the process of decoupling the positive and negative frequencies for the derivation of the conventional CMT in Eq. (S10), we exclude the contribution of QNM $n = 1$ for the direct path C. This contribution is $\int_{-2z_0/c_0}^{0} \widehat{K}_1(\tau)e^{-j\tau\omega_0}c_0/z_0 d\tau$, $\omega_0 = \widetilde{\omega}_{-1}$ and supposed to be of higher order since the operation frequency $\omega_0$ is far away from the complex frequency $\widetilde{\omega}_1$ of the QNM $n = 1$. Indeed, in the scenario here, it approaches to $(2 - 16/\pi^2)/\varepsilon_r$ when $\varepsilon_r \to \infty$.

In the weak-coupling regime where $\varepsilon_r \to \infty$ and when $\sigma = 0$ for the lossless screen, we can also expand the other system parameters in Eq. (S10). Specifically, we have $K = D = j\sqrt{\frac{2c_0}{-z_0}}\left[\frac{1}{\sqrt{\varepsilon_r}} + \frac{1}{2\varepsilon_r^{3/2}} + O\left(\frac{1}{\varepsilon_r^{5/2}}\right)\right]$ and $\Gamma = \frac{c_0}{(-z_0)\varepsilon_r}\left[1 + \frac{1}{3\varepsilon_r} + O\left(\frac{1}{\varepsilon_r^2}\right)\right]$. We can easily see that in the leading order the system parameters as derived satisfy all the requirements imposed by the time-reversal symmetry and energy conservation principles, which are $D^+D = 2\Gamma$, $K = D$ and $CD^* = -D$ [68]. Finally, we remark that the performance of the conventional CMT as Eq. (S10) will not be improved when we incorporate higher-order terms in system parameters, since in its derivation the crucial (rotating-wave) approximation has been already made by decoupling the positive and negative frequencies. Typically, the approximated equation under the constraints of general principles will behave relatively better.

## II. Formation of a QNM EP

We analyze the formation of the QNM EP $\hat{y}(\varepsilon_r^{EP}) = -jp, p > 0$ as seen in Fig. 1(c) of the main text. This EP arising as the merging of the $n = \pm 1$ QNMs is of the second order. Therefore, around the EP, we expand the complex frequency $\hat{y}_{\pm 1}(\varepsilon_r)$ of the QNMs as $\hat{y}_{\pm 1}(\varepsilon_r) = -jp \pm x_1\sqrt{\Delta\varepsilon} + x_2\Delta\varepsilon + \cdots, \Delta\varepsilon = \varepsilon_r - \varepsilon_r^{EP}$, and intend to determine the parameters p and $x_1$ explicitly. To this end, we substitute this ansatz for $\hat{y}_{\pm 1}(\varepsilon_r)$ into the transcendental equation $\tan(\tilde{k}_n z_0) = -jc_0\tilde{k}_n/\widetilde{\omega}_n$ where $\tilde{k}_n = \frac{\sqrt{\varepsilon_r}}{c_0}\widetilde{\omega}_n\sqrt{1 - j\sigma/(\varepsilon_0\varepsilon_r\widetilde{\omega}_n)}$ for the complex frequencies $\widetilde{\omega}_n$, or equivalently

$$\tan\left(\hat{y}_n\sqrt{1 + j\frac{\hat{\sigma}}{\hat{y}_n\sqrt{\varepsilon_r^{EP} + \Delta\varepsilon}}}\right) = -j\sqrt{1 + j\frac{\hat{\sigma}}{\hat{y}_n\sqrt{\varepsilon_r^{EP} + \Delta\varepsilon}}}\sqrt{\varepsilon_r^{EP} + \Delta\varepsilon} \qquad (S11)$$

where $\hat{y}_n = \widetilde{\omega}_n z_0\sqrt{\varepsilon_r}/c_0$, $z_0 = -d$ and $\hat{\sigma} = \sigma d\sqrt{\mu_0/\epsilon_0}$. Then, expanding Eq. (S11) order by order with respect to $\Delta\varepsilon$ and equating the resulting coefficient of each power of $\sqrt{\Delta\varepsilon}$ to zero, we obtain up to $O(\sqrt{\Delta\varepsilon})$

$$\tanh\left(p\sqrt{1 - \frac{\hat{\sigma}}{p\sqrt{\varepsilon_r^{EP}}}}\right) = \sqrt{\varepsilon_r^{EP}}\sqrt{1 - \frac{\hat{\sigma}}{p\sqrt{\varepsilon_r^{EP}}}} \qquad (S12)$$

and

$$\frac{-x_1\left[\sqrt{\varepsilon_r^{EP}}\hat{\sigma} + p(\hat{\sigma} - 2p\sqrt{\varepsilon_r^{EP}})\text{sech}^2\left(p\sqrt{1 - \frac{\hat{\sigma}}{p\sqrt{\varepsilon_r^{EP}}}}\right)\right]}{2p^2\sqrt{\varepsilon_r^{EP}}\sqrt{1 - \frac{\hat{\sigma}}{p\sqrt{\varepsilon_r^{EP}}}}} = 0. \qquad (S13)$$

The emergence of the EP requires that $x_1 \neq 0$, and then combining Eqs. (S12) and (S13) allows us to evaluate the EP $\hat{y}(\varepsilon_r^{EP}) = -jp$ with

$$p = \frac{(3\varepsilon_r^{EP} - 1)\hat{\sigma} + \sqrt{\hat{\sigma}}\sqrt{(\varepsilon_r^{EP})^2(\hat{\sigma} - 8) + \hat{\sigma} + 2\varepsilon_r^{EP}(4 + \hat{\sigma})}}{4(\varepsilon_r^{EP} - 1)\sqrt{\varepsilon_r^{EP}}}. \qquad (S14)$$

In turn, with a given $\hat{\sigma}$ value for the normalized conductivity, substituting the expression for p in Eq. (S14) into Eq. (S12) allows us to determine the parameter $\varepsilon_r^{EP}$. For example, in the case of the main text when $\hat{\sigma} = \hat{\sigma}^{opt} = 1.998$, we get $\varepsilon_r^{EP} \approx 2.1$ and correspondingly $p \approx 1.84$ using Eq. (S14). To obtain the expansion coefficient $x_1$ of $\hat{y}_{\pm 1}(\varepsilon_r)$ around the EP with respect to $\sqrt{\Delta\varepsilon}$, we need to resort to the next order $O(\Delta\varepsilon)$, which after considering Eqs. (S12) and (S13) can be simplified to be

$$\frac{j\left\{4p^3\left(p\sqrt{\varepsilon_r^{EP}} - \hat{\sigma}\right) + x_1^2\sqrt{\varepsilon_r^{EP}}\hat{\sigma}\left[\left(4p\sqrt{\varepsilon_r^{EP}} - \hat{\sigma}\right)(1 + \hat{\sigma}) - 4p^2\varepsilon_r^{EP}\right]\right\}}{4p^3\sqrt{\varepsilon_r^{EP}}\left(2p\sqrt{\varepsilon_r^{EP}} - \hat{\sigma}\right)\sqrt{1 - \frac{\hat{\sigma}}{p\sqrt{\varepsilon_r^{EP}}}}} = 0. \qquad (S15)$$

Interestingly, the expansion coefficient $x_2$ is absence in Eq. (S15) due to the relationships in Eqs. (S12) and (S13), which enables an analytical result for the expansion coefficient $x_1$:

$$x_1 = \left\{ \frac{4p^3\left(p\sqrt{\varepsilon_r^{EP}} - \hat{\sigma}\right)}{\sqrt{\varepsilon_r^{EP}}\hat{\sigma}\left[\left(\hat{\sigma} - 4p\sqrt{\varepsilon_r^{EP}}\right)(1+\hat{\sigma}) + 4p^2\varepsilon_r^{EP}\right]} \right\}^{1/2}. \tag{S16}$$

In the case when $\hat{\sigma} = \hat{\sigma}^{opt} = 1.998$, we have $x_1 \approx 1.53$.

## References

[66] Rigorously speaking, there exists one extra trivial QNM for our lossy absorber associated with the complex frequency $\widetilde{\omega}_0 = j\sigma/(\varepsilon_0\varepsilon_r)$ and wavenumber $\widetilde{k}_0 = 0$, which, however, should not be included in the QNM expansion since in this case the mode profile $|\widetilde{\Psi}_0(z)\rangle$ is a zero vector and non-normalizable, see Eq. (4) of the main text.

[67] H. A. Haus, *Waves and Fields in Optoelectronics* (Prentice-Hall, Englewood Cliffs, NJ, 1984).

[68] W. Suh, Z. Wang, and S. Fan, "Temporal coupled-mode theory and the presence of non-orthogonal modes in lossless multimode cavities," IEEE J. Quantum Electron. **40**, 1511 (2004).